\documentclass[twocolumn,showpacs]{revtex4}

\usepackage{graphicx}
\usepackage{amsmath}

\begin{document}

\title{Statefinder diagnostic for cosmology with the abnormally weighting energy hypothesis}

\author{Dao-jun Liu}\email{cufa3@shnu.edu.cn}\affiliation{Center for Astrophysics, Shanghai Normal University,
100 Guilin Road, Shanghai, 200234, China}
\author{Wei-zhong Liu}
\affiliation{Center for Astrophysics, Shanghai Normal University,
100 Guilin Road, Shanghai, 200234, China}
\date{\today}

\begin{abstract}
In this paper, we apply the statefinder diagnostic to the cosmology with the Abnormally Weighting Energy hypothesis (AWE cosmology), in which dark energy in the observational (ordinary matter) frame results from the violation of weak equivalence principle (WEP) by pressureless matter.  It is  found that there exist closed loops in the statefinder plane, which is an interesting characteristic of the evolution trajectories of statefinder parameters  and can be used to  distinguish AWE cosmology from the other cosmological models.
\end{abstract}


 \maketitle


Understanding the acceleration of the cosmic expansion is one of the deepest problems of modern cosmology and physics. In order to explain the acceleration, an unexpected  energy component of the cosmic budget, dark energy, is introduced by many cosmologists. Perhaps the simplest proposal is the Einstein's cosmological constant $\Lambda$ (vacuum energy), whose energy density remains constant with time. However, due to some conceptual problems  associated with the cosmological constant (for a review, see \cite{ccp}), a large variety of alternative possibilities have been explored. The most popular among them is quintessence scenario which uses a scalar field $\phi$ with a suitably chosen potential $V(\phi)$ so as to make the vacuum energy vary with time. Inclusion of a non-minimal coupling to gravity in quintessence models together with further generalization leads to models of dark energy in a scalar-tensor theory of gravity. Besides, some other models invoke unusual material in the universe such as Chaplygin gas, tachyon, phantom or k-essence (see, for a review, \cite{dde} and reference therein). The possibility that dark energy comes from the modifications  of four-dimensional general theory of relativity (GR) on large scales due to the presence of extra dimensions \cite{DGP} or other assumptions \cite{fR} has also been explored. A merit of these  models is the absence of matter violating the strong energy condition (SEC).

Recently, F\"{u}zfa and Alimi propose a completely new interpretation of dark energy that does also not require the violation of strong energy condition \cite{AWE}. They assume that dark energy does not couple to gravitation as usual matter and weights abnormally, \textit{i.e.}, violates the weak equivalence principle (WEP) on large scales. The abnormally weighting energy (AWE) hypothesis naturally derives from more general effective theories of gravitation motivated by string theory in which the couplings of the different matter fields to the dilaton are not universal in general (see \cite{AWE07} and the reference therein). In Ref.\cite{AWE07}, F\"{u}zfa and Alimi also applied the above  AWE hypothesis to a pressureless fluid to explain dark energy effects and further to consider a unified approach to dark energy and dark matter.

As so many dark energy models have been proposed, a discrimination between these rivals is needed. A new geometrical diagnostic, dubbed the statefinder pair $\{r, s\}$ is proposed by Sahni \textit{et al} \cite{statefinder}, where $r$ is only determined by the scalar factor $a$ and its derivatives with respect to the cosmic time $t$, just as the Hubble parameter $H$ and the deceleration parameter $q$, and $s$ is a simple combination of $r$ and $q$. The statefinder pair has been used to explor a series of dark energy and cosmological models, including $\Lambda$CDM, quintessence, coupled quintessence,  Chaplygin gas, holographic dark energy models, braneworld models, Cardassion models and so on \cite{SF03,Pavon,TJZhang}.

In this paper, we apply the statefinder diagnostic to the AWE cosmology. We find that there is a typical  characteristic of the evolution of statefinder parameters for the AWE cosmology that can be distinguished from the other cosmological models.

As is presented in Ref.\cite{AWE07}, in the AWE cosmology, the energy content of the universe is divided into three parts: a gravitational sector with metric field ($g_{\mu\nu}^{*}$ ) and  scalar field ($\phi$) components, a matter sector containing the usual fluids (baryons, photons, normally weighting dark matter if any, etc) and an abnormally weighting energy (AWE) sector. The normally and abnormally weighting matter are assumed to interact only through their gravitational influence without any direct interaction.
The corresponding action in the Einstein frame can be written as
\begin{eqnarray}\label{action}
S&=&\frac{1}{2\kappa_*}\int \sqrt{-g_*}d^4x\{R^*-2g_*^{\mu\nu}\partial_{\mu}\phi\partial_{\nu}\phi\}\nonumber\\
&+&S_m[\psi_m, A^2_m(\phi)g^*_{\mu\nu}]+S_{awe}[\psi_{awe}, A^2_{awe}(\phi)g^*_{\mu\nu}]
\end{eqnarray}
where $S_m$ is the action for the  matter sector with matter fields $\psi_m$, $S_{awe}$ is the action for AWE sector with fields $\psi_{awe}$, $R^*$ is the curvature scalar, $\kappa_{*}=8\pi G_*$ and $G_*$ is the 'bare' gravitational coupling constant. $A_{awe}(\phi)$ and $A_{m}(\phi)$ are the constitutive coupling functions to the metric $g_{\mu\nu}^*$ for the AWE and matter sectors respectively.

 Considering a flat Friedmann-Lemaitre-Robertson-Walker(FLRW) universe with metric
 \begin{equation}\label{line1}
 ds_*^2=-dt_*^2+a_*^2(t_*)dl_*^2,
 \end{equation}
 where $a_*(t_*)$ and $dl_*$  are the scale factor and Euclidean line element in the Einstein frame.
 The Friedmann equation derived from the action (\ref{action}) is
 \begin{equation}
 H_*^2=\left(\frac{1}{a_*}\frac{da_*}{dt_*}\right)^2
 =\frac{({d\phi}/{dt_*})^2}{3}+\frac{\kappa_*}{3}(\rho_m^*+\rho_{awe}^*)
 \end{equation}
 where $\rho_m^*$ and $\rho_{awe}^*$ are energy density of normally and abnormally weighting matter respectively.
Assuming further that both the matter sector and AWE sector are constituted by a pressureless fluid, one can obtain the evolution of $\rho_m^*$ and $\rho_{awe}^*$,
\begin{equation}
\rho_{m,awe}^*=A_{m,awe}(\phi)\frac{C_{m,awe}}{a_*^3},
\end{equation}
where $C_{m,awe}$ are two constants to be specified. Introducing a new variable $\lambda=\ln (a_*/a_*^i)$ where $a_*^i$ is a constant, the Klein-Gordon equation ruling the scalar field dynamics is reduced to be
\begin{equation}\label{KG1}
\frac{2\phi''}{3-\phi'^2}+\phi'+\frac{R_c\alpha_m(\phi)A_{m}(\phi)+\alpha_{awe}(\phi)A_{awe}(\phi)}{R_cA_{m}(\phi)+A_{awe}(\phi)}=0,
\end{equation}
where a prime denotes a derivative with respect to $\lambda$, the parameter $R_c= C_m/C_{awe}$ and the functions $\alpha_{m,awe}=d\ln(A_{m,awe}(\phi))/d\phi$.

 However, Einstein frame, in which the physical degrees of freedom are separated, is not correspond to a physically observable frame. Cosmology and more generally everyday physics are built upon observations based on "normal" matter which couples universally to a unique metric $g_{\mu\nu}$ and according to the AWE action (\ref{action}), $g_{\mu\nu}$ defines the observational frame through the following conformal transformation:
 \begin{equation}
 g_{\mu\nu}=A_m^2(\phi)g_{\mu\nu}^*.
 \end{equation}
 Therefore, the line element of FLRW metric (\ref{line1}) in the observational frame can be written down as
 \begin{equation}
 ds^2=-dt^2+a^2(t)dl^2,
 \end{equation}
where the scale factor $a(t)$ and the element of cosmic time read
\begin{equation}
a(t)=A_m(\phi)a_*(t_*)=e^{\lambda}A_m(\phi)a^i_*,\;\;dt=A_m(\phi)dt_*.
\end{equation}

Therefore, the Friedmann equation in the observational frame reads

\begin{eqnarray}\label{H21}
H^2\equiv\left(\frac{\dot{a}}{a}\right)^2
=\frac{8\pi G_*}{3}\frac{C_m}{a^3}\frac{A_m^2(\phi)\left(1+\frac{A_{awe}(\phi)}{A_m(\phi)}R_c^{-1}\right)}{\left(1-\alpha_m(\phi)\frac{d\phi}{dN}\right)^2-\frac{1}{3}\left(\frac{d\phi}{dN}\right)^2},
\end{eqnarray}
where  the overdot denotes the derivation with respect to the time $t$ and $N\equiv \ln(a/a^i)$ ( $a^i$ is value of the scale factor when $t=t_i$).
Further, the Friedmann equation (\ref{H21}) can be rewritten as
\begin{equation}
{H^2}={H^2_i}E^2(N)
\end{equation}
where
\begin{eqnarray}\label{E}
E^2(N)&=&Fe^{-3N}\frac{A_m^2(\phi)\left(1+\frac{A_{awe}(\phi)}{A_{m}(\phi)}R_c^{-1}\right)}
{\left(1-\alpha_m(\phi)\frac{d\phi}{dN}\right)^2-\frac{1}{3}\left(\frac{d\phi}{dN}\right)^2},
\end{eqnarray}
where the constant $F$
\begin{equation}
F=\frac{\left(1-\alpha_m(\phi_i)\frac{d\phi}{dN}|_i\right)^2-\frac{1}{3}\left(\frac{d\phi}{dN}|_i\right)^2}
{A_m^2(\phi_i)\left(1+\frac{A_{awe}(\phi_i)}{A_{m}(\phi_i)}R_c^{-1}\right)},
\end{equation}
where $\phi_i$ is the value of field $\phi$ at the moment $t=t_i$ and obviously, at this moment $E=E(0)=1$.

We can search for a translation of the AWE cosmology to usual dark energy cosmology.
In observational frame, one can search for an effective dark energy density parameter $\Omega_{DE}$ together with its effective equation of state $w$ in a spatially flat universe
\begin{equation}\label{E22}
E^2(N)=\Omega_M^{i}e^{-3N}+\Omega_{DE}(N),
\end{equation}
\begin{equation}\label{OmegaDE}
\Omega_{DE}(N)=(1-\Omega_M^{i})f_X(N),
\end{equation}
and
\begin{equation}\label{wDE}
w(N)=-1-\frac{1}{3}\frac{d\ln f_X(N)}{dN},
\end{equation}
where $\Omega_M^{i}$ is the total amount of effective dust matter energy density parameter at the time $t=t_i$.
Therefore, from Eq.(\ref{E}), we have
\begin{equation}
f_X(N)=\frac{e^{-3N}}{1-\Omega_M^{i}}\left(F\frac{A_m^2(\phi)\left(1+\frac{A_{awe}(\phi)}{A_{m}(\phi)}R_c^{-1}\right)}
{\left(1-\alpha_m(\phi)\frac{d\phi}{dN}\right)^2-\frac{1}{3}\left(\frac{d\phi}{dN}\right)^2}-\Omega_M^{i}\right).
\end{equation}

Let us choose the coupling functions $A_m{\phi}$ and $A_{awe}(\phi)$ as:
\begin{equation}\label{model}
  A_m(\phi) = \exp\left(k_m \frac{\phi^2}{2}\right), \;\;  A_{awe}(\phi) = \exp\left(k_{awe} \frac{\phi^2}{2}\right),
\end{equation}
where $k_m$ and $k_{awe}$ are two arbitrarily constant. Therefore,
\begin{equation}
  \alpha_m(\phi) = k_m \phi, \;\;\; \alpha_{awe}(\phi) = k_{awe} \phi.
\end{equation}
Note that we assume here that $k_m\neq k_{awe}$, otherwise, it is corresponds to the case of an ordinary scalar-tensor theory.
Meanwhile,  the Klein-Gordon equation for $\phi$ in the observational frame (\ref{KG1}) reduced to be
\begin{widetext}
\begin{equation}\label{KG3}
\frac{2\left(\frac{d^2\phi}{dN^2}+k_m \left(\frac{d\phi}{dN}\right)^3\right)}
{\left(1-k_m \phi\frac{d\phi}{dN}\right)^3}
+\left(3-\frac{\left(\frac{d\phi}{dN}\right)^2}{\left(1-k_m \phi\frac{d\phi}{dN}\right)^2}\right)\left(\frac{\frac{d\phi}{dN}}{1-k_m \phi\frac{d\phi}{dN}}
+k_m \phi+\frac{(k_{awe}-k_m) \phi}{1+R_c \exp\left((k_m-k_{awe}) \frac{\phi^2}{2}\right)}\right)=0.
\end{equation}
\end{widetext}

It is not difficult to find that if $R_c+k_{awe}/k_m>0$, Eq.(\ref{KG3}) have only one limited critical point $(\phi_c,\frac{d\phi}{dN}|_c)=(0,0)$, which is an unstable saddle point. However, in the case of $R_c+k_{awe}/k_m<0$, their exists two critical points $(0,0)$ and $\left(\sqrt{\frac{2}{k_m-k_{awe}}\ln \left(-R_c^{-1}\frac{k_{awe}}{k_m}\right)},0\right)$. It is shown, by performing a linear stability analysis, that the critical point (0,0) is unstable  in this case, but $\left(\sqrt{\frac{2}{k_m-k_{awe}}\ln \left(-R_c^{-1}\frac{k_{awe}}{k_m}\right)},0\right)$ is stable.  Note that $R_c$ denoting the ratio of energy density of ordinary  and AWE matter is certainly positive and we also assume that $k_m >0$ in Eq.(\ref{model}). Therefore, if we set a condition that $k_{awe}<-R_c k_m $, an attractor solution of scalar field $\phi$ is expected. From now on, we only deal with the models with this prior condition.

The traditional geometrical diagnostics, i.e., the Hubble parameter $H$ and the deceleration parameter $q\equiv {-\ddot{a}a}/{\dot{a}^2}$, are two good choices to describe the expansion state of our universe but they can not characterize the cosmological models uniquely, because a quite number of models may just correspond to the same current value of $H$ and $q$. Fortunately, as is shown in many literatures, the statefinder pair $\{r, s\}$ which is also a geometrical diagnostic, is able to distinguish a series of cosmological models successfully.

The statefinder pair $\{r,s\}$ defines two new cosmological parameters in addition to $H$ and $q$:
\begin{equation}
r\equiv \frac{\dddot{a}}{aH^3},\;\;\;\;s\equiv \frac{r-1}{3(q-1/2)}.
\end{equation}
As an important function, the statefinder can allow us to differentiate between a given dark energy model and the simplest of all models,  i.e., the cosmological constant $\Lambda$. For the $\Lambda$CDM model, the statefinder diagnostic pair $\{r,s\}$ takes the constant value $\{1,0\}$, and for the SCDM model, $\{1,1\}$.
The statefinders $r$ can be easily expressed in terms of the Hubble parameter $H(z)$ and its derivatives as follows:
\begin{equation}
r(x)=1-2\frac{H'}{H}x+\left[\frac{H''}{H}+\left(\frac{H'}{H}\right)^2\right]x^2,
\end{equation}
where the variable $x=1+z$,
$q(x)=\frac{H'}{H}x-1$ and $H'$ is the derivative of $H$ with respect to the redshift $z$, and immediately $s$ is also a function of $x$.
We can now use this tool to explore the evolutionary trajectories of the universe governed by AWE cosmology.

\begin{center}
\begin{figure}
  \includegraphics[width=5.5cm]{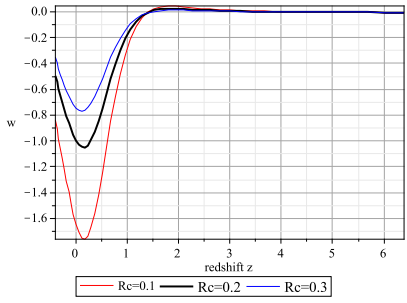}\\
  \caption{The evolution of equation-of-state $w$ for the effective dark energy in the observational frame, where we choose the parameters $k_m=1$,$k_{awe}=-10$ and $R_c=0.1,0.2,0.3$, respectively.}\label{w-z}
\end{figure}
\end{center}

\begin{center}
\begin{figure}
  \includegraphics[width=5.5cm]{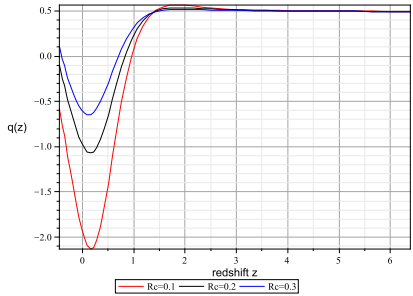}\\
  \caption{The evolution of deceleration parameter $q(z)$ in the observational frame, where we choose the parameters $k_m=1$,$k_{awe}=-10$ and $R_c=0.1,0.2,0.3$, respectively.}\label{q-z}
\end{figure}
\end{center}

In FIG. \ref{w-z}, we show the evolutionary trajectories of equation of state of effective dark energy in observational frame for model (\ref{model}) with different parameters. For some parameters that is chosen, the equation-of-state parameter $w$ is able to cross the cosmological constant divide $w=-1$ between phantom and quintessence.From FIG. \ref{q-z}, we find that at high redshifts the standard matter-dominated  cosmology is recovered and at low redshifts the universe become accelerating and dark energy dominated as expected. The transition from deceleration to acceleration occurs roughly at $z\approx 1$.

\begin{center}
\begin{figure}
  \includegraphics[width=5.5cm]{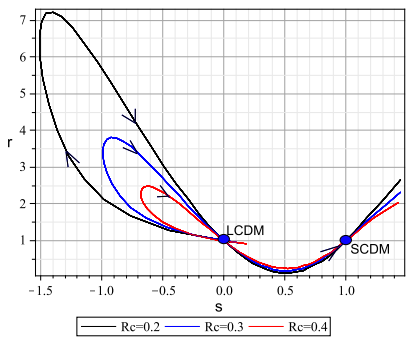}\\
  \caption{Trajectories in the statefinder plane $\{r,s\}$ for the model (\ref{model}), where the parameters of the model is chosen as $k_m=1$,$k_{awe}=-10$ and $R_c=0.2,0.3,0.4$, respectively. Two circles at the point $\{1,0\}$ and $\{1,1\}$ in the plane denote the fixed statefinder pairs of LCDM and SCDM model, respectively. The arrows show the direction of the time evolution.}\label{r-s}
\end{figure}
\end{center}

\begin{center}
\begin{figure}
  \includegraphics[width=5.5cm]{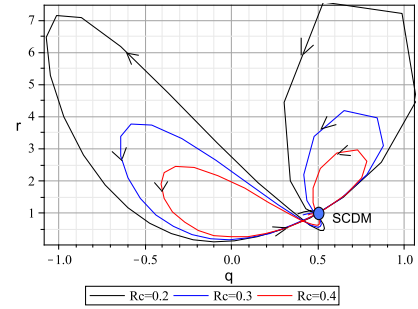}\\
  \caption{Trajectories in the statefinder plane $\{r,q\}$ for the model (\ref{model}), where the parameters of the model is chosen as $k_m=1$,$k_{awe}=-10$ and $R_c=0.2,0.3,0.4$, respectively. The arrows show the direction of the time evolution.}\label{r-q}
\end{figure}
\end{center}

The time evolution trajectories of statefinder pairs $\{r,s\}$ and $\{r,q\}$ for model (\ref{model}) are shown in FIG. \ref{r-s} and FIG. \ref{r-q}, respectively.  The most interesting characteristic of the trajectories is that there is a loop in the plane. Along the time evolution, after passing through the $\Lambda$CDM fixed point $\{r=1,s=0\}$, the statefinder pairs is now going along with a loop in the plane, and at some time in the future they will pass through the $\Lambda$CDM fixed point again. After that, they will go towards  the SCDM fixed point $\{r=1,s=1\}$. This character of the trajectories is significantly different from those of the other cosmological models, such as  quintessence,  Chaplygin gas,  brane world ( see, for example, \cite{SF03}), phantom \cite{sfphantom}, Cardassian \cite{TJZhang}, holographic dark energy \cite{xzhang05a} agegraphic dark energy models \cite{caihao}, and so on. From FIG. \ref{r-q}, it is obviously that the acceleration of the universe in this model is a transient phenomenon. In the past, the standard matter-dominated universe is simply mimicked, but in the future, although a SCDM state is an attractor, the universe will go through a series of states, which are different from that of SCDM but decelerating, before the attractor is finally reached. It is worth noting that a class of braneworld models, called "disappearing dark energy" (DDE) \cite{sahni}, in which the current acceleration of the universe is also a transient phase and  there exists closed loop in the $\{r,q\}$ plane \cite{SF03}, but there is no closed loop  which contains the $\Lambda$CDM fixed point $\{r=1,s=0\}$ in the $\{r,s\}$ plane as that of the model studied in this work.


In summary,  we investigate the cosmology with the hypothesis of abnormally weighting energy by using the statefinder diagnostic. The statefinder diagnosis provides a useful tool to break the possible degeneracy of different
cosmological models by constructing the parameters $\{r, s\}$ or $\{r, q\}$ using the higher derivative of
the scale factor. It is  found that the trajectories of the statefinder pairs $\{r,s\}$ and $\{r,q\}$ of AWE cosmology in the statfinder plane have a typical characteristic which is distinguished from other cosmological models. We hope that the future high-precision observations offer more accurate data to determine the model parameters more precisely, rule out some models  and  consequently  shed light on the nature of dark energy.

\begin{acknowledgments}
This work is supported in part by National Natural Science Foundation of
China under Grant No. 10503002 and Shanghai
Commission of Science and technology under Grant No. 06QA14039.
\end{acknowledgments}

\end{document}